\documentclass[aps,prl,reprint,superscriptaddress,nofootinbib]{revtex4-1}
\usepackage{amsmath,amssymb}
\usepackage{graphicx}
\usepackage[caption=false]{subfig}
\usepackage[export]{adjustbox}
\usepackage{color}
\usepackage{bm}

\usepackage{accents}

\usepackage{subfiles}
\usepackage{enumerate}

\allowdisplaybreaks

\newcommand{\nd}{\mathrm{d}}

\begin{document}

\title{Symbol letters of Feynman integrals from Gram determinants}

\author{Xuhang Jiang}
\email{xhjiang@itp.ac.cn}
\affiliation{CAS Key Laboratory of Theoretical Physics, Institute of Theoretical Physics, Chinese Academy of Sciences, Beijing 100190, China}

\author{Jiahao Liu}
\email{liujiahao@itp.ac.cn}
\affiliation{CAS Key Laboratory of Theoretical Physics, Institute of Theoretical Physics, Chinese Academy of Sciences, Beijing 100190, China}
\affiliation{School of Physical Sciences, University of Chinese Academy of Sciences, No.19A Yuquan Road, Beijing 100049, China}

\author{Xiaofeng Xu}
\email{xiaxu@uni-mainz.de}
\affiliation{PRISMA \hspace{-0.15cm}$^+$ Cluster of Excellence, Johannes Gutenberg University, 55099 Mainz, Germany}

\author{Li Lin Yang}
\email{yanglilin@zju.edu.cn}
\affiliation{Zhejiang Institute of Modern Physics, School of Physics, Zhejiang University, Hangzhou 310027, China}


\begin{abstract}
Symbol letters are crucial for analytically calculating Feynman integrals in terms of iterated integrals. We present a novel method to construct the symbol letters for a given integral family without prior knowledge of the canonical differential equations. We provide a program package implementing our algorithm, and demonstrate its effectiveness in solving non-trivial problems with multiple loops and legs. Using our method, we successfully bootstrap the canonical differential equations for a two-loop five-point family with two external masses and for a three-loop four-point family with two external masses, which were previously unknown in the literature. We anticipate that our method can be applied to a wide range of cutting-edge calculations in the future.
\end{abstract}


\maketitle

\section{Introduction}

The language of symbols has emerged as one of the most powerful tools for analytically calculating Feynman integrals since its initial introduction to the community \cite{Goncharov:2009lql, Goncharov:2010jf}. It captures the beautiful mathematical structures underlying Chen's iterated integrals \cite{Chen:1977oja}, which naturally arise as solutions to the canonical differential equations (CDEs) satisfied by a broad range of Feynman integrals \cite{Henn:2013pwa}. In this Letter, we focus on iterated integrals of the $\nd\log$-form, expressed as
\begin{equation}
    f = \int \nd\log\alpha_n \cdots \int \nd\log\alpha_2 \int \nd\log\alpha_1 \,,
\end{equation}
where the integration is path-ordered along a contour from the boundary point to the target point within the space of kinematic parameters. The symbol of $f$ is defined as $\mathcal{S}(f) = \alpha_1 \otimes \alpha_2 \otimes \cdots \otimes \alpha_n$, where the $\alpha_i$'s are functions of the kinematic parameters and are referred to as ``symbol letters''. The knowledge of the symbol $\mathcal{S}(f)$ completely determines the function $f$ up to a boundary condition.

Traditionally, obtaining symbol letters for a family of Feynman integrals involves several steps. First, a system of differential equations satisfied by the master integrals in the family is derived using integration-by-parts (IBP) \cite{Chetyrkin:1981qh} reduction techniques. These differential equations are then transformed into the canonical form $\partial_{y_i}\vec{f} = \epsilon \bm{B}_i(\bm{y}) \vec{f}$, where $\vec{f}$ represents the vector of master integrals, $\bm{y}$ denotes the set of kinematic parameters (with $y_i$ being one of them), and $\bm{B}_i$ is a matrix that depends on $\bm{y}$. The subsequent step involves combining the partial derivatives into a total derivative, resulting in the CDEs of the form $\nd \vec{f} = \epsilon \, \nd\bm{A}(\bm{y}) \vec{f}$. The symbol letters can then be identified by examining the entries in the matrix $\bm{A}$.

In cutting-edge multi-loop problems involving a large number of scales, the conventional procedure outlined above faces considerable challenges. Symbolic IBP reduction can be computationally demanding, and determining the CDEs and combining partial derivatives into total derivatives becomes increasingly difficult as the number of loops and kinematic parameters grows. 

The complexity of these problems necessitates the development of innovative approaches and techniques to obtain symbol letters, circumventing the difficulties in the conventional procedure. Having symbol letters at hand offers several benefits. First, the knowledge of symbol letters can be used as input to find the canonical bases, e.g., with the \texttt{INITIAL} algorithm \cite{Dlapa:2020cwj}. It is then straightforward to bootstrap the matrix $\bm{A}$ in the CDEs. It also allows bootstrapping analytic expressions of Feynman integrals and even the full scattering amplitudes in terms of iterated integrals \cite{Dixon:2011pw, Dixon:2013eka, Drummond:2014ffa, Chicherin:2017dob, Caron-Huot:2019vjl}. Therefore, it is highly desirable to have an algorithmic method to directly obtain the symbol letters of generic multi-loop Feynman integrals.

In recent years, numerous efforts have been made towards this goal. The structure of one-loop symbol letters has been fully understood from various perspectives \cite{Spradlin:2011wp, Arkani-Hamed:2017ahv, Abreu:2017mtm, Caron-Huot:2021xqj, Chen:2022fyw}. At higher loops, attempts have been made utilizing the concepts of Landau singularities \cite{Dennen:2015bet, Prlina:2018ukf, Mizera:2021icv, Hannesdottir:2021kpd, Lippstreu:2022bib, Dlapa:2023cvx, Fevola:2023kaw, Fevola:2023fzn, Helmer:2024wax}, intersection theory \cite{Chen:2023kgw} and the Schubert analysis \cite{Yang:2022gko, He:2022ctv, Morales:2022csr, He:2022tph, He:2023umf, He:2024fij}. The Landau analysis provides a powerful tool to obtain the rational letters, as well as the singularities of the algebraic ones. However, one cannot directly obtain the explicit form of algebraic letters from those information. The method of Schubert analysis is another promising approach that works for algebraic letters as well, but at the moment is not fully automated.

Besides the above methods that aim at obtaining letters for a whole integral family, there are also complementary approaches that aim at solving the symbols for a single integral. Examples include the \texttt{HyperInt} algorithm~\cite{Panzer:2014caa} and the recently proposed geometric method~\cite{Gong:2022erh}. These methods have their limitations and do not work for generic integrals at the moment.

In this Letter, we present a streamlined approach to construct candidates of symbol letters, including both rational and algebraic ones, for a given integral family. We provide a proof-of-concept \texttt{Mathematica} package \cite{github} to demonstrate the remarkable capabilities of our method in a couple of multi-loop multi-leg examples. Our algorithm contains two main steps: the identification of rational letters from leading singularities (LS) in the Baikov representation (BR) \cite{Baikov:1996iu, Lee:2010wea}, and the construction of algebraic ones from well-educated ansatz. These two steps will be described in the following two sections.

\section{Setup and rational letters}

\begin{figure}[tb]
    \subfloat{\includegraphics[width=0.2\textwidth,valign=c]{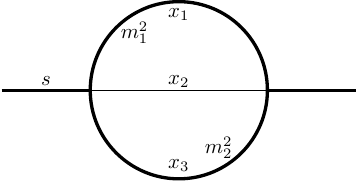}}
    \quad
    \subfloat{\includegraphics[width=0.2\textwidth,valign=c]{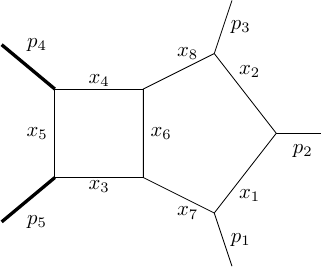}}
    \\
    \subfloat{\includegraphics[width=0.2\textwidth,valign=c]{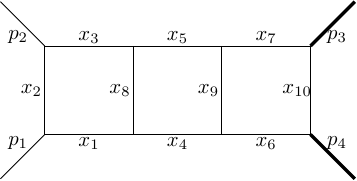}}
    \caption{\label{fig:diagrams}Some of the examples discussed in this Letter.}
\end{figure}

In this section, we lay out the theoretical foundations of our algorithm, and discuss our approach of identifying the LS that serve as candidates for the rational letters. For pedagogical purposes, we illustrate the relevant concepts using the two-mass sunrise family that has been well-studied in the literature \cite{Czyz:2002re, He:2023umf}. This family is shown as the first diagram in Fig.~\ref{fig:diagrams}. 

We begin with the BRs for an integral family, characterized by a collection of Baikov variables $\bm{x} \equiv \{x_i\}$ corresponding to propagator denominators and irreducible scalar products (ISPs). For the sunrise family, the Baikov variables are defined as:
\begin{align}
    x_1 &= l_1^2 - m_1^2 \,,\; x_2 = (l_1-l_2)^2 \,,\; x_3 = (l_2-p)^2 - m_2^2 \,, \nonumber
    \\
    x_4 &= l_2^2 \,, \; x_5 = (l_1 - p)^2 \,,
\end{align}
where $l_1$ and $l_2$ represent loop momenta, and $p$ denotes the external momentum. The integrals in this family depend on the kinematic parameters $m_1^2$, $m_2^2$ and $s \equiv p^2$. We specifically focus on the top-sector labeled by $\{1,2,3\}$, where $x_1,x_2,x_3$ are propagator denominators and $x_4,x_5$ are ISPs. Additionally, we consider all sub-sectors within this top-sector. The relevant integrals can be expressed in the BRs as follows:
\begin{equation}
    \label{eq:baikov_rep}
    \int_{\mathcal{C}} u(\bm{x}) \, \prod_i \frac{\nd x_i}{x_i^{a_i}} \,, \quad u(\bm{x}) = \prod_j \left[ P_j(\bm{x}) \right]^{\alpha_j + \beta_j\epsilon} \,,
\end{equation}
where the $P_j$'s are polynomials of the Baikov variables and the kinematic parameters derived from the Gram determinants (Grams for short) of loop and external momenta. The Gram of two momentum sequences $\bm{k}$ and $\bm{q}$ is defined as $G(\{\bm{k}\},\{\bm{q}\}) \equiv \det(k_i \cdot q_j)$, and we will also employ the shorthand notation $G(\bm{k}) \equiv G(\{\bm{k}\},\{\bm{k}\})$.

\begin{figure}[tb]
    \includegraphics[width=0.3\textwidth]{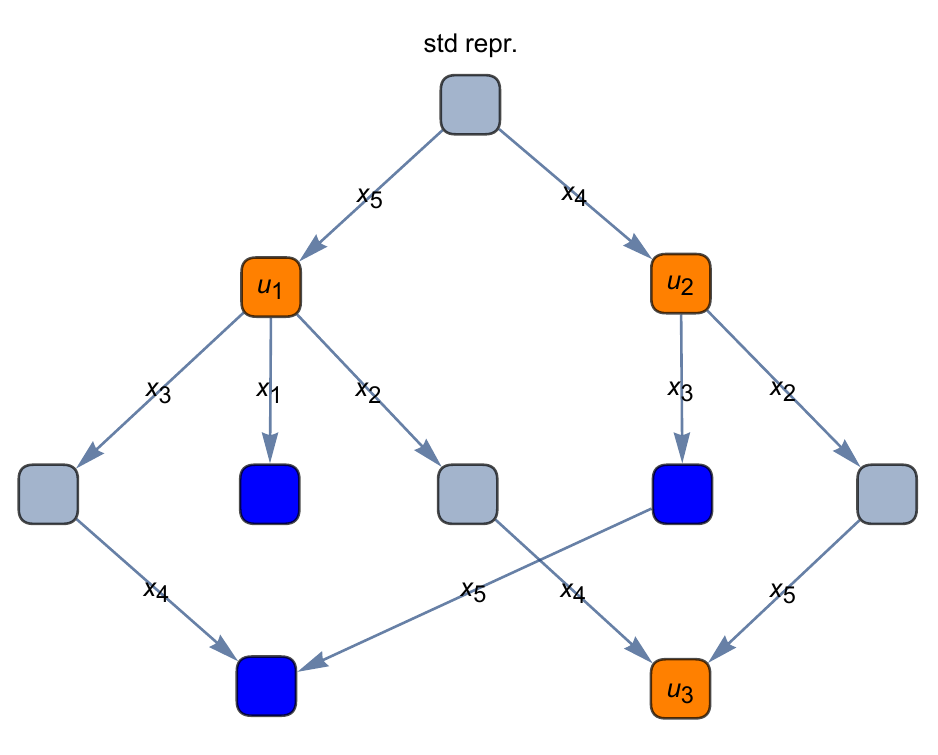}
    \caption{\label{fig:baikov_recursive}The recursive structure of Baikov representations for the sunrise family.}
\end{figure}

According to the findings in \cite{Jiang:2023qnl}, an integral family is not associated with just a single BR, but rather a tree of BRs that form a recursive structure. Each BR is defined by a specific $u$-function in Eq.~\eqref{eq:baikov_rep}. The so-called standard BR, which includes all Baikov variables in $u$, serves as the topmost block in the recursive structure. { It takes the form: 
\begin{multline}\label{eq:baikovFI}
    I_{\text{std}}=\frac{\pi^{(L-N)/2}\det(A^a_{ij})}{\prod_{i=1}^{L}\Gamma\left(\frac{d-K+i}{2}\right)} \left[ G(p_1,\ldots,p_E) \right]^{-(d-E-1)/2}
    \\
    \times \int\frac{\nd x_1\cdots\nd x_N}{x_1^{a_1}\cdots x_N^{a_N}} \left[ P^{L}_{N}(x_1,\cdots,x_N) \right]^{(d-K-1)/2} \, ,
\end{multline}
where $E$ is the number of independent external legs, $L$ is the number of loops, and $K\equiv L+E$. The Baikov polynomial $P^{L}_{N}$ is given by
\begin{equation}
    P^{L}_{N}(x_1,\ldots,x_N)=G(q_1,q_2,\ldots,q_K) \, ,
\end{equation}
where $\{q_{1},\ldots,q_{K}\}$ is the set of all loop and independent external momenta. The scalar products $q_{i}\cdot q_{j}$ can then be expressed as linear combinations of the Baikov variables.}
The standard BR is the super-BR of all other BRs. Given a BR, one can derive its sub-BRs by sequentially integrating out variables. {It is proven in \cite{Jiang:2023qnl} that such variables can always be found. They are quadratic in the polynomial and can be integrated out using the general formula
\begin{equation}\label{eq:recursionformula}
    \begin{aligned}
    &\int_{c_1}^{c_2}z^{n} \left[-A(z-c_1)(z-c_2)\right]^{\gamma}\mathrm{d}z=(-A)^{\gamma} \, (c_2-c_1)^{1+2\gamma}\\
    &\frac{\Gamma(1\!+\!\gamma)^2}{\Gamma(2(1\!+\!\gamma))}\!\!\left(\!\!\frac{c_1+c_2}{2}\!\!\right)^{n}
    \!\!\!{}_{2}F_{1}\left(-\frac{n}{2},\!\frac{1\!-\!n}{2},\!\frac{3}{2}\!+\!\gamma;\left(\!\frac{c_1\!-\!c_2}{c_1\!+\!c_2}\!\right)^2\!\right) .
    \end{aligned}
\end{equation}
When $n\ge 0$, the hypergeometric function ${}_{2}F_{1}$ truncates to a polynomial. In particular, when $n=0,1$, it equals to 1. This integration can be performed recursively due to the structure of Gram determinants, thus all the BRs form a tree-and-net structure.} In the case of the $\{1,2,3\}$ top-sector and its sub-sectors in the sunrise family, the recursive structure is illustrated in Fig.~\ref{fig:baikov_recursive}. The arrows in the diagram are labeled with the variables to be integrated out. The blue blocks represent zero-BRs, where all integrals are scaleless and vanish. Each sector can possess several BRs. For instance, the $\{1,2,3\}$ sector can be expressed in terms of the $u_1$, $u_2$ and the standard BRs. Among them, $u_1$ and $u_2$ are the lowest ones in their branches, i.e., they don't have descendants that can be used to represent the sector $\{1,2,3\}$. We will refer to them as the ``minimal BRs'' for this sector. This concept will play an important role in our construction. It is typical that a genuine multi-loop sector has more than one minimal BRs. 
A ``contra-example'' is the $\{1,3\}$ sector, which can be represented by the $u_3$-BR and all its super-BRs as depicted in Fig.~\ref{fig:baikov_recursive}. It can be seen that $u_3$ is the only minimal BR for this sector. This reflects the fact that integrals in the $\{1,3\}$ sector are products of one-loop integrals.

With the above setup, we can now proceed to identify the rational letters, which is done by recursively applying the following procedure for each non-zero sector.
We select all the minimal BRs associated with that sector, and perform the maximal cut on each of them, which involves setting all propagator denominators to zero in the corresponding $u$-function. This leads to a $u$-function depending only on ISPs. After that, we further localize the Grams at the singular points of this $u$-function. We will refer to this operation as ``maximal-localization'', which fixes the values of all Baikov variables including both propagator denominators and ISPs.
This operation then leads to the LS, which can be written in terms of polynomials raised to some (integer or half-integer) powers. The LS with half-integer powers will be referred to as ``algebraic LS''. These polynomials serve as candidates for rational symbol letters. However, there is often redundancy when multiple minimal BRs exist for a sector, and we apply a further criterion to get rid of the spurious singularities. This criterion is based on the number of master integrals in a sector (i.e., the dimension of the linear space of integrals): at a true singularity this dimension should always decrease, irrespective of the representation used.
It is worth noting that the above analysis of LS is conceptually equivalent to the Landau analysis in \cite{Dennen:2015bet, Prlina:2018ukf, Mizera:2021icv, Hannesdottir:2021kpd, Lippstreu:2022bib, Dlapa:2023cvx, Fevola:2023kaw, Fevola:2023fzn, Helmer:2024wax}, albeit in a different representation. In particular, the dimension criterion is similar to the criterion based on the Euler characteristic employed in~\cite{Fevola:2023kaw, Fevola:2023fzn}.

As a pedagogical example of our method, consider the $\{1,2,3\}$ sector in the sunrise family. After applying the maximal cut $x_1,x_2,x_3 \to 0$, the $u$-function becomes:
\begin{align}
    \tilde{u}_1(x_4) &= \left[ \tilde{G}(p) \tilde{G}(l_2) \right]^\epsilon \left[ \tilde{G}(l_1,l_2) \tilde{G}(l_2,p) \right]^{-1/2-\epsilon} \,, \nonumber
    \\
    \tilde{G}(p) &= s \,, \quad \tilde{G}(l_2,p) = -\lambda(x_4,s,m_2^2)/4 \,, \nonumber
    \\
    \tilde{G}(l_2) &= x_4 \,, \quad \tilde{G}(l_1,l_2) = -(x_4-m_1^2)^2/4 \,,
\end{align}
where we use a tilde to denote the function after maximal cut, and $\lambda(x,y,z) \equiv x^2+y^2+z^2-2xy-2yz-2zx$ is the K\"all\'en function. Here, we have set the spacetime dimension to $d=2-2\epsilon$. To analyze the singularities, it is advantageous to promote the space of $\{x_i\}$ to a projective space, enabling a unified treatment of the singularity at infinity. In this specific case, we introduce the projective coordinates $[x_4 : x_0]$ and homogenize $\tilde{u}_1$ by multiplying each term by an appropriate power of $x_0$. The LS can then be obtained by localizing the factors of Grams in $\tilde{u}_1$ at the rational singular points: $[0 : 1]$, $[m_1^2 : 1]$, and $[1 : 0]$. For the $\{1,2,3\}$ sector, we get four maximally-localized Grams or rational letter candidates: $\{s, m_1^2, m_2^2, \lambda(m_1^2,s,m_2^2)\}$. Note that we have removed two spurious letters $s-m_1^2$ and $s-m_2^2$ using the dimension criterion. The $u_3$-BR for the $\{1,3\}$ sector does not provide new candidates.

\section{Bootstrapping algebraic letters}

The information obtained in the previous step (including the rational singular points and the associated rational letters) serve as crucial inputs for determining the algebraic letters. Inspired by the generic results of one-loop letters \cite{Abreu:2017mtm, Chen:2022fyw, Jiang:2023qnl} and the $\nd\log$-integrands~\cite{Chen:2020uyk, Chen:2022lzr}, we take the ansatz for the algebraic letters as $W(P,Q) \equiv (P+\sqrt{Q})/(P-\sqrt{Q})$, where $P$ and $Q$ are polynomials of kinematic variables arising from the maximal localization of certain Grams in the $u$-functions. We use $\hat{G}$ to denote the maximal-localization of a Gram $G$ at one of the rational singular points determined in the previous section. The polynomials in the ansatz are then given by $P = \hat{B}$, $Q = \hat{A} \hat{C}$ or $Q = -\hat{D} \hat{E}$. The five Grams $A,B,C,D,E$ are defined as
\begin{align}
    B &= G(\{\bm{k},q_{i}\},\{\bm{k},q_{j}\}) \,, \; A = G(\bm{k},q_{i}) \,, \nonumber
    \\
    C &= G(\bm{k},q_{j}) \,, \; D = G(\bm{k}) \,, \; E = G(\bm{k},q_{i},q_{j}) \,,
\end{align}
where $\bm{k}$ represents a sequence of momenta, while $q_i$ and $q_j$ denote two momenta not present in $\bm{k}$. {Note that $\bm{k}$ $q_{i}$ and $q_{j}$ can be loop momenta. The five Grams are not chosen arbitrarily, but are taken from the recursive structure of the BRs. Their forms are determined once we specify how the variables are integrated out from the topmost BR.} For convenience, we also allow $\bm{k}$ to be an empty sequence, and define $G() = 1$. These Grams are linked by the identity $B^2 + DE = AC$ \cite{Dlapa:2021qsl, Chen:2022lzr}, which leads to
\begin{align}
    \partial_B \log W(B,AC) \Big|_{AC} &= \frac{2\sqrt{AC}}{DE} \,, \nonumber
    \\
    \partial_B \log W(B,AC) \Big|_{DE} &= \frac{2}{\sqrt{AC}} \,,
    \label{eq:deriv_dlog}
\end{align}
where the subscript $AC$ (or $DE$) indicates that $AC$ (or $DE$) should be treated as a constant when taking the partial derivative with respect to $B$. Two additional identities can be obtained by interchanging $AC$ with $-DE$. We now discuss further criteria to be applied on $W(\hat{B},\hat{A}\hat{C})$, while those for $W(\hat{B},-\hat{D}\hat{E})$ are similar.

\paragraph{Criterion 1: compatibility with powers in $u$ and with the algebraic LS.} We categorize all the Grams in the family based on their powers $\alpha+\beta\epsilon$ in the $u$-functions. We define the sets $\mathcal{G}_1$ containing those with $\alpha \in \mathbb{Z}$ and $\mathcal{G}_2$ containing those with $\alpha+1/2 \in \mathbb{Z}$. We require that either $A,C \in \mathcal{G}_2$ or $D,E \in \mathcal{G}_1$. Furthermore, we require that $\sqrt{\hat{A}\hat{C}}$ is a product of the algebraic LS obtained in the previous section. We also demand that $\hat{A}\hat{C}$ is not a perfect square, otherwise it would result in a rational letter. 

\paragraph{Criterion 2: compatibility with the recursive structure.} We require that either $A,C$ or $D,E$ are linked by the one-directional arrows in the graph of the recursive structure. In other words, if two Grams $G_1,G_2$ are linked, then $G_1$ should appear in either a super- or a sub-BR of where $G_2$ appears.

\paragraph{Criterion 3: compatibility with rational letters.} We require that both $\hat{A}\hat{C}$ and $\hat{D}\hat{E}=-(\hat{B}+\sqrt{\hat{A}\hat{C}})(\hat{B}-\sqrt{\hat{A}\hat{C}})$ can be factorized into products of the rational letters obtained previously. This condition arises from the fact that all singularities within the family should already be contained in the rational letters.

We again use the sunrise family as an example, where the categories $\mathcal{G}_2$ and $\mathcal{G}_1$ are
\begin{align}
\mathcal{G}_2 &= \{ G(), G(l_1,p), G(l_2,p), G(l_1,l_2), G(l_1-p,l_2-p)\} \,, \nonumber \\
\mathcal{G}_1 &= \{ G(), G(p), G(l_1-p), G(l_2-p), \nonumber \\
&\hspace{10em} G(l_1),G(l_2),G(l_1,l_2,p) \} \,.
\end{align}
Applying the aforementioned criteria allows us to identify several candidates in the form $W(\hat{B},\hat{A}\hat{C})$ or $W(\hat{B},-\hat{D}\hat{E})$, two of which are independent. They are given by
\begin{align}
    DE &= G() \, G(l_2,p) \,, \quad [x_4 : x_0] \to [m_1^2 : 1] \,, \nonumber
    \\
    DE &= G() \, G(l_1,p) \,, \quad [x_5 : x_0] \to [m_2^2 : 1] \,,
\end{align}
where the arrows specify the rational singular points in the projective spaces where the Grams should be localized after maximal cut. These exactly correspond to the two independent algebraic letters in the CDEs.

Having outlined our algorithm in the last two sections, we now turn to present some intriguing results obtained using its automated implementation.

\section{Examples}

\noindent \textbf{Validation using existing results.} We first use several existing results in the literature to validate our method, including a range of two-loop five-point and three-loop four-point integral families \cite{Gehrmann:2015bfy, Gehrmann:2018yef, Chicherin:2020oor, Abreu:2020jxa, Abreu:2021smk, Canko:2021xmn, Badger:2022hno, Henn:2023vbd}.

\paragraph{Two-loop pentabox with one external mass.} There are three configurations with the external massive leg inserted at different positions. They are denoted as mzz, zmz and zzz in \cite{Abreu:2020jxa}. For these three families, we get 27, 30 and 31 rational letters, and 11, 18 and 18 algebraic letters, respectively. We have analytically compared them with the results in \cite{Abreu:2020jxa} and find complete agreements.

\paragraph{Two-loop pentabox with internal mass.} We compare our results for two pentabox families denoted as $\mathrm{PB}_{A}$ and $\mathrm{PB}_{C}$ in \cite{Badger:2024fgb}. For $\mathrm{PB}_{A}$, we obtain 33 rational and 38 algebraic letters; and for $\mathrm{PB}_{C}$, we obtain 35 rational and 44 algebraic letters. All the letters agree analytically with those in \cite{Badger:2024fgb}.

\paragraph{Three-loop ladders and tennis court with one external mass.} We compare our results for several three-loop families denoted as $A$, $E_1$, $E_2$, $B_1$ and $B_2$ in \cite{Henn:2023vbd}. For the planar families $A, E_1, E_2$ (which has been studied in \cite{DiVita:2014pza,Canko:2021xmn}), we obtain 7 rational letters for each, which agree perfectly with the known results. However, there are a few algebraic letters that survive our criteria, but do not appear in the CDEs. These should be regarded as spurious. We have also applied our algorithm to the non-planar families $B_1$ and $B_2$, and obtain 11 and 9 rational letters. They fully contain the results of \cite{Henn:2023vbd}, but involve two spurious ones for each family (and again, there are a few spurious algebraic letters).

\noindent{\textbf{New results.}} We now showcase the remarkable capabilities of our method by presenting results for several exceptionally intricate integral families that have not been previously documented in the literature. These integral families hold significance not only in academic contexts but also in addressing cutting-edge phenomenological problems.

\paragraph{Two-loop pentabox with two external masses.} This is shown as the second diagram in Fig.~\ref{fig:diagrams} where $p_4$ and $p_5$ have different masses.
It is relevant for the production processes of multiple gauge bosons at hadron colliders. By running \texttt{Kira} \cite{Klappert:2020nbg} numerically, we identify a total of 127 master integrals within this family. To establish the canonical basis $\vec{f}$, we adopt the approach of \cite{Chen:2020uyk, Chen:2022lzr}, while incorporating the previously obtained results for certain sub-sectors \cite{Henn:2014lfa, Dlapa:2021qsl}. We predict 43 rational and 61 algebraic letters $W_i$ ($i=1,\cdots,104$). The remaining unknowns in the CDEs $\nd\vec{f} = \epsilon \sum_i \bm{A}_i \left( \nd\log W_i \right) \vec{f}$ are then the coefficient matrices $\bm{A}_i$ composed of rational numbers. We determine these rational numbers by choosing several sets of numeric values for the kinematic variables. This completes the bootstrap of the CDEs. The details about the bootstrap procedure and the explicit results are collected in the Appendix.







\paragraph{Three-loop triple-box with masses.} This is shown as the third diagram in Fig.~\ref{fig:diagrams}, which are of great relevance for numerous scattering processes that involve the Higgs boson, electroweak gauge bosons and top quarks at the three-loop level. We consider the specific configuration where $p_3$ and $p_4$ have the same mass $m$. Our method predicts 9 rational and 2 algebraic letters. During the bootstrap procedure, we find that one rational letter is spurious and do not appear in the CDEs.

We have also applied our method to other configurations in the tripe-box families, such as: $p_3$ and $p_4$ having different masses; $x_1$, $x_2$, $x_3$ and $x_8$ having the same mass. We have attached the results for these cases as a supplemental material. We leave the bootstrap of the corresponding CDEs for future work.

\section{Summary and outlook}

In this Letter, we present an algorithmic approach for constructing both rational and algebraic symbol letters for a given integral family that admits $\nd\log$-type CDEs. Our method does not require prior knowledge of the CDEs, thus bypassing challenging steps such as symbolic IBP reduction and the construction of canonical bases. We identify possible rational letters from the singular structure of the Baikov representations under maximal cut, and select candidates for algebraic letters as combinations of Gram determinants based on criteria derived from the recursive structure. We provide a proof-of-concept program package that automates our algorithm, demonstrating its effectiveness in solving non-trivial problems with multiple loops and external legs. These include both planar and non-planar families at two and three loops with up to five external legs.

The symbol letters obtained through our method can be used to bootstrap the CDEs for the corresponding integral families. This can be effectively accomplished with numeric IBP reduction and $\nd\log$-integrand construction. In particular, we successfully obtain the CDEs for a two-loop five-point family with two external masses and for a three-loop four-point family with two external masses. Both results were previously unknown in the literature. It is then easy to write the solutions as iterated integrals, which can be further converted to multiple polylogarithms (MPLs) \cite{Goncharov:1998kja, Goncharov:2001iea, Brown:2009qja} or evaluated using series expansion along a contour \cite{Moriello:2019yhu, Hidding:2020ytt}. We anticipate that our method can be applied to a wide range of similar calculations in the future.

One of the key elements in our approach is the identification of rational singular points, which correspond to ``pinched'' common roots of certain Gram determinants expressed as polynomials of the Baikov variables. This concept is closely related to Landau singularities, commonly studied in the Feynman parameter representation \cite{Dennen:2015bet, Prlina:2018ukf, Mizera:2021icv, Hannesdottir:2021kpd, Lippstreu:2022bib, Dlapa:2023cvx, Fevola:2023kaw, Fevola:2023fzn, Helmer:2024wax}. 
Our algorithm to identify the leading singularities (candidates for rational letters) provides a systematic approach to calculate all Landau loci in the Baikov representations, for MPL-type integral families. One can verify that the rational singular points obtained in our approach are indeed related to the solutions of Landau equations in Baikov representations~\cite{Caron-Huot:2024brh}. It will be interesting to extend such analysis to more complicated integral families, e.g, those involving elliptic curves or beyond.

Our method can be applied to non-planar topologies as well. For non-planar topologies, a complication is that there are usually more spurious letters appearing in our predictions. One way to reduce the number of spurious letters is to choose the ISPs appropriately, such that all propagator variables appear at most quadratically in the $u$-function of the standard representation. This can often be achieved by choosing the ISPs to be linear in loop momenta. Such choices simplify the recursive structure of BRs and also simplify the calculations. We elaborate more on these details in an upcoming article.

There can also be more complicated forms for the algebraic letters than our ansatz. We have found that in certain non-planar families (e.g., the double-pentagon family studied in \cite{Abreu:2023rco}), some algebraic letters fail to be expressed in terms of maximally-localized Grams. There are also situations in which CDEs exist, but the symbol letters cannot be expressed in the $\nd\log$ form \cite{FebresCordero:2023gjh}. These letters typically involve LS with nested square roots that are not in our ansatz. It remains an open question to determine whether these symbol letters can still be derived from specific Gram determinants. This topic is left for future investigations.

Finally, it would be intriguing to explore the relationship between our method and the approach based on Schubert problems \cite{Yang:2022gko, He:2022ctv, Morales:2022csr, He:2022tph, He:2023umf}. We provide a brief discussion on this topic in the Appendix.

\vspace{1.5ex}
\begin{acknowledgments}
We would like to thank Qinglin Yang and Song He for their valuable discussions and collaboration on relevant works.
This work was supported in part by the National Natural Science Foundation of China under Grant No. 12375097, 12147103, 12047503 and 12225510, and the Fundamental Research Funds for the Central Universities. This research is supported by the Cluster of Excellence PRISMA$^+$, ``Precision Physics, Fundamental Interactions and Structure of Matter" (EXC 2118/1) within the German Excellence Strategy (project ID 390831469).This research has received funding from the European Research Council (ERC) under the European Union’s Horizon 2022 Research and Innovation Program (ERC Advanced Grant agreement No.101097780, EFT4jets). Views and opinions expressed are however those of the authors only and do not necessarily reflect those of the European Union or the European Research Council Executive Agency. Neither the European Union nor the granting authority can be held responsible for them. 
\end{acknowledgments}





\onecolumngrid

\appendix

\section{A. Relation to the Schubert approach}

It is interesting to investigate the relation between our approach and the Schubert approach of \cite{Yang:2022gko, He:2022ctv, Morales:2022csr, He:2022tph, He:2023umf}. The relation is particularly evident when performing the Schubert analysis in the embedding formalism \cite{Dirac:1936fq, Weinberg:2010fx, Simmons-Duffin:2012juh}. For simplicity, we will focus on one-loop integrals, while higher-loop ones can be studied loop-by-loop. In the embedding formalism, each vector in the $d$-dimensional Minkowski space is embedded into a $(d+2)$-dimensional projective space with two time-like directions. Consider a propagator denominator $x_i = (l-p_i)^2 - m_i^2$, where $l$ denotes the loop momentum and $p_i$ denotes a combination of external momenta. The embedded vectors are given by
\begin{equation}
l \hookrightarrow Y = (l^\mu; -l^2, 1) \,, \; p_i \hookrightarrow P_i = (p_i^\mu; -p_i^2 + m_i^2, 1) \,,
\end{equation}
where the last two components are given in the light-cone coordinates. The scalar products are given by $(Y,Y) = 0$, $(P_i,P_i) = m_i^2$ and $(Y,P_i) = -x_i/2$. It is useful to introduce the infinity vector $I_\infty = (0^\mu, -1, 0)$, and the propagator denominator can then be written as $x_i = (Y,P_i)/(Y,I_\infty)$. The possible singular points of the integrals are then determined by the conditions $(Y,P_i) \to 0$ ($x_i \to 0$) and $(Y,I_\infty) \to 0$ ($x_0 \to \infty$).

A Schubert problem amounts to finding the solutions $Y$ for a set of equations $\{(Y,X_i) = 0\}$, where $X_i$ can be one of the external momenta $\{P_j\}$ or $I_\infty$. In general, each Schubert problem has two solutions labelled as $Y_{\pm}$. We use superscripts to label different Schubert problems. For example, $Y^1_{\pm}$ and $Y^2_{\pm}$ are solutions to two different Schubert problems, respectively. The Schubert approach then suggests that the cross ratio
\begin{equation}
\frac{(Y^1_{+},Y^2_{+}) (Y^1_{-},Y^2_{-})}{(Y^1_{+},Y^2_{-})(Y^1_{-},Y^2_{+})} \,,
\end{equation}
provides a candidate for the symbol letters. In the following, we show how these cross ratios are related to the symbol letters written in terms of Gram determinants obtained using our method.

We consider two Schubert problems associated with a one-loop box integral and a one-loop triangle integral near $d=4$ dimensions. The external momenta in the propagators of the box integral are $P_0,P_1,P_2,P_3$ in the embedding space, and we assign $X_i=P_i$ for $i=0,1,2,3$ and $X_4=I_\infty$. In the triangle integral, $P_3$ is absent. The two Schubert problems are given by
\begin{align}
(Y^1,X_0) &= (Y^1,X_1) = (Y^1,X_2) = (Y^1,X_3) = 0 \,, \nonumber
\\
(Y^2,X_0) &= (Y^2,X_1) = (Y^2,X_2) = (Y^2,X_4) = 0 \,.
\end{align}
To solve the two sets of equations, we introduce an auxiliary independent vector $X_5$ in the $(d+2)$-dimensional space. We further introduce a set of dual vectors $\tilde{X}_i$ for $i=0,\ldots,5$ satisfying $(X_i,\tilde{X}_j) = \delta_{ij}$. The dual vector $\tilde{X}_i$ can be naturally expressed in terms of the six-dimensional completely antisymmetric tensor contracted with the five vectors $\{X_k\, (k \neq i)\}$, with suitable normalizations.
The solutions to the first Schubert problem can then be parameterized as
\begin{equation}
Y^1_{\pm} = \tilde{X}_4 + c_{\text{box}}^{\pm} \tilde{X}_5 \,,
\end{equation}
where the coefficient $c_{\text{box}}$ is determined by $Y^2=0$:
\begin{equation}
c_{\text{box}}^{\pm} = - \frac{G(\{X_0,X_1,X_2,X_3,X_4\},\{X_0,X_1,X_2,X_3,X_5\}) \pm \sqrt{G(X_0,X_1,X_2,X_3)G(X_0,X_1,X_2,X_3,X_4,X_5)}}{G(X_0,X_1,X_2,X_3,X_4)} \,.
\end{equation}
For the second set of equations, the solutions are given by
\begin{equation}
Y^2_{\pm} = \tilde{X}_3 + c_{\text{tri}}^{\pm} \tilde{X}_5 \,,
\end{equation}
where
\begin{equation}
c_{\text{tri}}^{\pm} = - \frac{G(\{X_0,X_1,X_2,X_3,X_4\},\{X_0,X_1,X_2,X_4,X_5\}) \pm \sqrt{G(X_0,X_1,X_2,X_4)G(X_0,X_1,X_2,X_3,X_4,X_5)}}{G(X_0,X_1,X_2,X_3,X_4)} \,.
\end{equation}
The cross ratio is then
\begin{equation}
\frac{(Y^1_{+},Y^2_{+}) (Y^1_{-},Y^2_{-})}{(Y^1_{+},Y^2_{-})(Y^1_{-},Y^2_{+})} = \left( \frac{G(\{X_0,X_1,X_2,X_3\},\{X_0,X_1,X_2,X_4\}) + \sqrt{G(X_0,X_1,X_2,X_3)G(X_0,X_1,X_2,X_4)}}{G(\{X_0,X_1,X_2,X_3\},\{X_0,X_1,X_2,X_4\}) - \sqrt{G(X_0,X_1,X_2,X_3)G(X_0,X_1,X_2,X_4)}} \right)^2 .
\end{equation}
It can be observed that the arbitrary auxiliary vector $X_5$ drops out in the cross ratio. The expression of the cross ratio in terms of Gram determinants exactly corresponds to the result obtained through our method.

\section{B. Results for a two-loop pentabox family with two different masses}

\begin{figure}[t!]
    \includegraphics[width=0.6\textwidth]{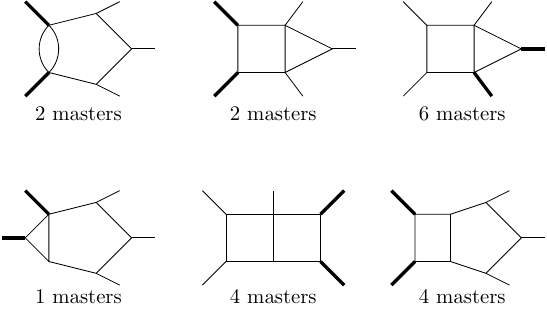}
    \caption{Genuine configurations of planar two-loop five-point integrals with two adjacent massive external legs.}
    \label{fig:pentabox_two_masses}
\end{figure}

In this appendix, we present the detailed results for the two-loop pentabox family with two massive external legs. This top topology of this family is shown as the last diagram in Fig.~\ref{fig:pentabox_two_masses}. The Baikov variables are defined by
\begin{equation}
  \begin{aligned}
  &x_1=(l_1+p_1)^2, \; x_2=(l_1+p_1+p_2)^2, \; x_3=l_2^2, \; x_4=(l_2+p_1+p_2+p_3)^2, \\
  &x_5=(l_2+p_1+p_2+p_3+p_4)^2, \; x_6=(l_1-l_2)^2, \; x_7=l_1^2 , \; x_8=(l_1+p_1+p_2+p_3)^2, \\
  &x_9=(l_1+p_1+p_2+p_3+p_4)^2, \; x_{10}=(l_2+p_1)^2, \; x_{11}=(l_2+p_1+p_2)^2 \,,
  \end{aligned}
\end{equation}
where $x_9, x_{10} $ and $x_{11}$ are ISPs. The kinematic variables are
\begin{equation}
  \begin{aligned}
  &p_{1}^2=p_{2}^2=p_{3}^2=0, \; p_{4}^2=m_4^2, \; p_{5}^{2}=m_5^2,  \\
  &(p_1+p_2)^2=s_{12}, \; (p_2+p_3)^2=s_{23}, \; (p_3+p_4)^2=s_{34}, \; (p_4+p_5)^2=s_{45}, \; (p_5+p_1)^2=s_{15} \,.
  \end{aligned}
\end{equation}
By assigning numeric values for the above kinematic variables, we can easily run \texttt{Kira} and determine the number of master integrals in each sector of this family. There are 127 master integrals in total.

\subsection{Construction of the canonical basis}

The integrals for lower-point sub-sectors (i.e., those with less than five external legs) are already known in the literature \cite{Dlapa:2021qsl, Henn:2014lfa}. Therefore we only need to construct the canonical integrals for the genuine five-point configurations. These are shown in Fig.~\ref{fig:pentabox_two_masses}. In the following, we demonstrate the construction using several non-trivial examples.

We start from the box-triangle sector corresponding to the second diagram in Fig.~\ref{fig:pentabox_two_masses}. One of its minimal Baikov representation can be written as 
\begin{equation}\label{eq:BaikovPentatri}
	\int \bigg[ \prod_{i \neq 7,8,9} \nd x_{i} \bigg]  P_{1}^{-1-\epsilon} P_{2}^{-\frac{1}{2}+\epsilon} P_{3}^{-\epsilon} \frac{1}{x_1^{a_{1}}x_2^{a_{2}}\ldots x_{6}^{a_{6}}} \, ,
\end{equation}
where we have suppressed an irrelevant overall constant prefactor. The Baikov polynomials are given by
\begin{align}
   P_1(x_3,x_4,x_5,x_{10},x_{11}) &= G(l_2, p_1,p_2,p_3,p_4)\,, \nonumber
   \\
   P_2(x_{10},x_{11}) &= (x_{10}-x_{11})^2 \,, \nonumber
   \\
   P_3(x_1,x_2,x_6,x_{10},x_{11}) &= G(l_1+p_1,l_2+p_1,p_2)\,.
\end{align}
We need to construct two independent canonical integrals in this sector. We employ the $\nd\log$-integrand construction in \cite{Chen:2020uyk, Chen:2022lzr}. Motivated by the form of $P_2(x_{10},x_{11})$, we introduce the variable change $x'_{10}=x_{10}-x_{11}, \, x'_{11}=x_{11}$. One can then construct two $\nd\log$-integrals. The first one is simply
\begin{equation}
	\int P_1^{-\epsilon} P_2^{\epsilon} P_3^{-\epsilon} \, \nd \log x'_{10} \wedge \nd \log P_1 \wedge \nd \log x_1 \cdots \wedge \nd \log x_6 \, .
\end{equation}
The second $\nd\log$-integrand involves the discriminant $Q$ of the polynomial $P_1$ with respect to the variable $x'_{11}$:
\begin{multline}
		\int P_1^{-\epsilon} P_2^{\epsilon} P_3^{-\epsilon} \, \frac{\sqrt{Q}}{P_1} \nd x'_{11} \wedge \frac{\sqrt{Q(x_3=0)}}{x_3 \sqrt{Q}} \nd x_{3} \wedge \frac{\sqrt{Q(x_3,x_4=0)}}{x_4 \sqrt{Q(x_3=0)}} \nd x_{4} \wedge \frac{\sqrt{Q(x_3,x_4,x_5=0)}}{x_5 \sqrt{Q(x_3,x_4=0)}} \nd x_{5}\\
		 \wedge \frac{\sqrt{Q(x_3,x_4,x_5,x'_{10}=0)}}{x'_{10} \sqrt{Q(x_3,x_4,x_5=0)}} \nd x'_{10} \wedge \nd \log x_1 \wedge \nd \log x_2 \wedge \nd \log x_6 \, .
\end{multline}
The above two $\nd\log$-integrands in the Baikov representation can be easily converted to Feynman integrals. 

We now turn to the top sector. One of its minimal Baikov representations can be written as
\begin{equation}
  \int \bigg[\prod_{i\ne 10,11}\nd x_{i} \bigg] P_{1}^{\epsilon} P_{2}^{-1/2-\epsilon} P_{3}^{-1-\epsilon} \frac{1}{x_1^{a_1}x_2^{a_2}\ldots x_{8}^{a_{8}}} \, ,
\end{equation}
where again we have suppressed a constant prefactor, and the Baikov polynomials are
\begin{equation}
  \begin{aligned}
    P_{1}(x_7,x_8,x_9) &= -G(l_1,p_1+p_2+p_3,p_4) \,, \\
    P_{2}(x_3,x_4,x_5,x_6,x_7,x_8,x_9) &= G(l_1,l_2,p_1+p_2+p_3,p_4) \,, \\
    P_{3}(x_1,x_2,x_7,x_8,x_9) &= G(l_1,p_1,p_2+p_3,p_3,p_4) \,.
  \end{aligned}
\end{equation}
There are four master integrals in this sector. It is easy to construct one $\nd\log$-integral by noticing that $P_{1}$ is a codimension-one minor of $P_{2}$:
\begin{equation}
    f_1 = \int P_1^{\epsilon} P_2^{-\epsilon} P_3^{-\epsilon} \,  \frac{\sqrt{\lambda(m_4^2,m_5^2,s_{45})}}{P_{1}(x_7,x_8,x_9)\sqrt{P_{2}(x_3,x_4,x_5,x_6,x_7,x_8,x_9)}} \frac{\partial P_2}{\partial x_5} \, \nd x_{9} \bigwedge_{i=1}^8 d\log x_i \,.
\end{equation}
where $\lambda$ denotes the K\"all\'en function: $\lambda(x,y,z) = x^2 + y^2 + z^2 - 2xy - 2yz - 2zx$. The above integrand is equivalent to the following Feynman integral in $6-2\epsilon$ dimension:
\begin{equation}
  f_1 = \sqrt{\lambda(m_4^2,m_5^2,s_{45})} \sqrt{G(p_1,p_2,p_3,p_4)} \, I^{(6-2\epsilon)}(1,1,1,1,2,1,1,1,0,0,0) \,,
\end{equation}
which can be easily converted to Feynman integrals in $4-2\epsilon$ dimension.

It is non-trivial to construct the remaining three canonical integrals completely. However, the construction is straightforward under maximal cut. This would give rise to pre-canonical integrals whose differential equations are $\epsilon$-factorized within the top sector, but have non-factorized $\epsilon$-dependence in sub-sectors. One can then bring the sub-sector dependence to the $\epsilon$-form by systematically adding suitable combinations of sub-sector integrals. In the above top sector, besides $f_1$ that is already canonical, we can further construct three pre-canonical integrals $\tilde{f}_2$, $\tilde{f}_3$ and $\tilde{f}_4$. As an example, $\tilde{f}_2$ can be written as
\begin{equation}
\tilde{f}_2 = \epsilon^4 s_{12} s_{23} s_{45} \, I(1,1,1,1,1,1,1,1,-1,0,0) \,,
\end{equation}
whose derivatives have non-factorized $\epsilon$-dependence on two sub-sector canonical integrals:
\begin{align}
    f_5 &= \epsilon^4 s_{12} (m_5^2 s_{23}-s_{15}s_{45}) \, I(1, 1, 1, 1, 1, 1, 1, 0, 0, 0, 0) \,, \nonumber
    \\
    f_6 &= \epsilon^4 s_{23} (m_4^2 s_{12}-s_{34}s_{45}) \, I(1, 1, 1, 1, 1, 1, 0, 1, 0, 0, 0) \,.
\end{align}
We can write the differential equations of $f_2$ as
\begin{equation}
    \frac{\partial}{\partial y_j} \tilde{f}_2 = A_{5,j}(\epsilon,\bm{y}) f_5 + A_{6,j}(\epsilon,\bm{y}) f_6 + \epsilon \sum_{i \neq 5,6} A_{i,j}(\bm{y}) \accentset{\scriptscriptstyle(\sim)}{f}_{i} \,,
\end{equation}
where we use $\bm{y}=\{s_{12},s_{23},s_{34},s_{45},s_{15},m_4^2,m_5^2\}$ to collectively denote the kinematic variables. Considering the mass dimensions and transcendental weights of relevant integrals, we can make the following ansatz: 
\begin{equation}
	f_2 = \tilde{f}_2 + \epsilon^4 \left(\sum_{i,j,k=1}^7 c^{(1)}_{ijk} y_i y_j y_k \right) I(1, 1, 1, 1, 1, 1, 1, 0, 0, 0, 0) + \epsilon^4 \left(\sum_{i,j,k=1}^7 c^{(2)}_{ijk} y_i y_j y_k \right) I(1, 1, 1, 1, 1, 1, 0, 1, 0, 0, 0) \,,
\end{equation}
where the coefficients $ c^{(1)}_{ijk} $ and $ c^{(2)}_{ijk} $ are rational numbers. We require that the differential equation of $f_2$ is $\epsilon$-factorized. By computing the differential equation at several kinematic points, we can determine the coefficients and arrive at
\begin{equation}
	f_2 = \tilde{f}_2 - \epsilon^4  s_{45} \left[ s_{23} s_{34} \, I(1, 1, 1, 1, 1, 1, 0, 1, 0, 0, 0) + s_{12} s_{15} I(1, 1, 1, 1, 1, 1, 1, 0, 0, 0, 0) \right] .
\end{equation}
Through a similar procedure, we can get two further canonical integrals $f_3$ and $f_4$. This completes the construction of the whole canonical basis.

\subsection{Bootstrap the CDEs from symbol letters}

After having the canonical basis $\vec{f}=\{f_1,\cdots,f_{127}\}$, we can use the symbol alphabet $\{\alpha_{1},\cdots,\alpha_{104}\}$ predicted by our algorithm to bootstrap the CDEs. We write the ansatz for the CDEs as
\begin{equation}
	\nd \vec{f} = \epsilon \sum_{i=1}^{104} \bm{B}_{i} \left( \nd\log\alpha_{i} \right) \vec{f} \,,
\end{equation}
where $\bm{B}_i$ are $127 \times 127$ matrices of rational numbers. Choosing a set of numeric values for the kinematic variables, we can get a system of linear equations for the entries of $\bm{B}_i$. With at most 104 (the number of letters) different sets of numeric values, all the entries can be completely fixed, if there is no additional symbol letters beyond our predictions. In practice, we find that 43 rational letters and 61 algebraic letters obtained through our method are exactly those appearing in the CDEs. We have attached the results for the canonical basis and the CDEs as an electronic file in the supplemental materials.



\bibliography{references_inspire,
references_inspire_app,
references_local}


\end{document}